\documentclass[superscriptaddress,prl,twocolumn,amsmath,amssymb]{revtex4-1}
\usepackage{graphicx}
\usepackage{eurosym}
\usepackage{amsmath}
\usepackage{amssymb}
\usepackage{dcolumn}
\usepackage{bm}
\usepackage{subfigure}
\usepackage{hyperref}
\usepackage[final]{pdfpages}
\graphicspath{{../images/}}
\usepackage{soul}

\def\avg#1{\mathinner{\langle{#1}\rangle}}
\def\bra#1{\mathinner{\langle{#1}|}}
\def\ket#1{\mathinner{|{#1}\rangle}}
\newcommand{\braket}[2]{\langle #1|#2\rangle}

\newcommand\h{{\cal H}}

\newcommand{\ignore}[1]{}

\newcommand{\be}{\begin{equation}}
\newcommand{\ee}{\end{equation}}
\newcommand{\ba}{\begin{eqnarray}}
\newcommand{\ea}{\end{eqnarray}}

\usepackage{tabularx}
\makeatletter
\def\hlinewd#1{%
\noalign{\ifnum0=`}\fi\hrule \@height #1 %
\futurelet\reserved@a\@xhline}
\makeatother

\begin{document}
\title{
A variational eigenvalue solver on a quantum processor
}

\author{Alberto Peruzzo}
\thanks{These authors contributed equally to this work.}
\affiliation{Centre for Quantum Photonics, H.H.Wills Physics Laboratory \& Department of Electrical and Electronic Engineering, University of Bristol, Bristol BS8 1UB, UK}

\author{Jarrod McClean}
\thanks{These authors contributed equally to this work.}
\affiliation{Department of Chemistry and Chemical Biology, Harvard University, Cambridge MA, 02138}

\author{Peter Shadbolt}
\affiliation{Centre for Quantum Photonics, H.H.Wills Physics Laboratory \& Department of Electrical and Electronic Engineering, University of Bristol, Bristol BS8 1UB, UK}

\author{Man-Hong Yung}
\affiliation{Department of Chemistry and Chemical Biology, Harvard University, Cambridge MA, 02138}
\affiliation{Center for Quantum Information, Institute for Interdisciplinary
Information Sciences, Tsinghua University, Beijing, 100084, P. R.
China}

\author{Xiao-Qi Zhou}
\affiliation{Centre for Quantum Photonics, H.H.Wills Physics Laboratory \& Department of Electrical and Electronic Engineering, University of Bristol, Bristol BS8 1UB, UK}

\author{Peter J. Love}
\affiliation{Department of Physics, Haverford College, Haverford, PA 19041, USA}

\author{Al\'an Aspuru-Guzik}
\affiliation{Department of Chemistry and Chemical Biology, Harvard University, Cambridge MA, 02138}

\author{Jeremy L. O'Brien}
\affiliation{Centre for Quantum Photonics, H.H.Wills Physics Laboratory \& Department of Electrical and Electronic Engineering, University of Bristol, Bristol BS8 1UB, UK}

\begin{abstract}
\noindent Quantum computers promise to efficiently solve important problems that are intractable on a conventional computer. 
For quantum systems, where the dimension of the problem space grows exponentially, finding the eigenvalues of certain operators is one such intractable problem and remains a fundamental challenge. The quantum phase estimation algorithm can efficiently find the eigenvalue of a given eigenvector but requires fully coherent evolution. 
We present an alternative approach that greatly reduces the requirements for coherent evolution and we combine this method with a new approach to state preparation based on ans\"atze and classical optimization. We have implemented the algorithm by combining a small-scale photonic quantum processor with a conventional computer. We experimentally demonstrate the feasibility of this approach with an example from quantum chemistry---calculating the ground state molecular energy for He--H+, to within chemical accuracy.
The proposed approach, by drastically reducing the coherence time requirements, enhances the potential of the quantum resources available today and in the near future.
\end{abstract}
\maketitle
\noindent In chemistry, the properties of atoms and molecules can be determined by solving the Schr\"odinger equation. However, because the dimension of the problem grows exponentially with the size of the physical system under consideration, exact treatment of these problems remains classically infeasible for compounds with more than 2--3 atoms~\cite{Thogersen2004}. Many approximate methods~\cite{Helgaker2002} have been developed to treat these systems, but efficient exact methods for large chemical problems remain out of reach for classical computers. Beyond chemistry, the solution of large eigenvalue problems~\cite{Saad:1992} would have applications ranging from determining the results of internet search engines~\cite{Page:1999} to designing new materials and drugs~\cite{Golub:2000jy}.

Recent developments in the field of quantum computation offer a way forward for efficient solutions of many instances of large eigenvalue problems which are classically intractable~\cite{Nielsen:2007vn, kitaev:1996, Griffiths:1996, Neven:2008up, Harrow:2009gx, Berry:2010tp, Garnerone:2012}.
Quantum approaches to finding eigenvalues have previously relied on the quantum phase estimation (QPE) algorithm. The QPE algorithm offers an exponential speedup over classical methods and requires a number of quantum operations $O(1/p)$ to obtain an estimate with precision $p$~\cite{Abrams1997,Abrams1999,Aspuru:2005,Lanyon:2010,Whitfield:2011,Walther:2012}. In the standard formulation of QPE, one assumes the eigenvector $\ket{\psi}$ of a Hermitian operator $\h$ is given as input and the problem is to determine the corresponding eigenvalue $\lambda$. The time the quantum computer must remain coherent is determined by the necessity of $O(1/p)$ successive applications of $e^{-i \h t}$, each of which can require on the order of millions or billions of quantum gates for practical applications~\cite{Whitfield:2011,Jones:2012}, as compared to the tens to hundreds of gates achievable in the short term.  

Here we introduce and experimentally demonstrate an alternative to QPE that significantly reduces the requirements for coherent evolution. 
We have developed a reconfigurable quantum processing unit (QPU), which efficiently calculates the expectation value of a Hamiltonian ($\h$), providing an exponential speedup over conventional methods. The QPU is combined with an optimization algorithm run on a classical processing unit (CPU), which variationally computes the eigenvalues and eigenvectors of $\h$. By using a variational algorithm, this approach reduces the requirement for coherent evolution of the quantum state, making more efficient use of quantum resources, and may offer an alternative route to practical quantum-enhanced 
computation. 

\begin{figure*}[t]
\centering
\includegraphics[width = 12cm]{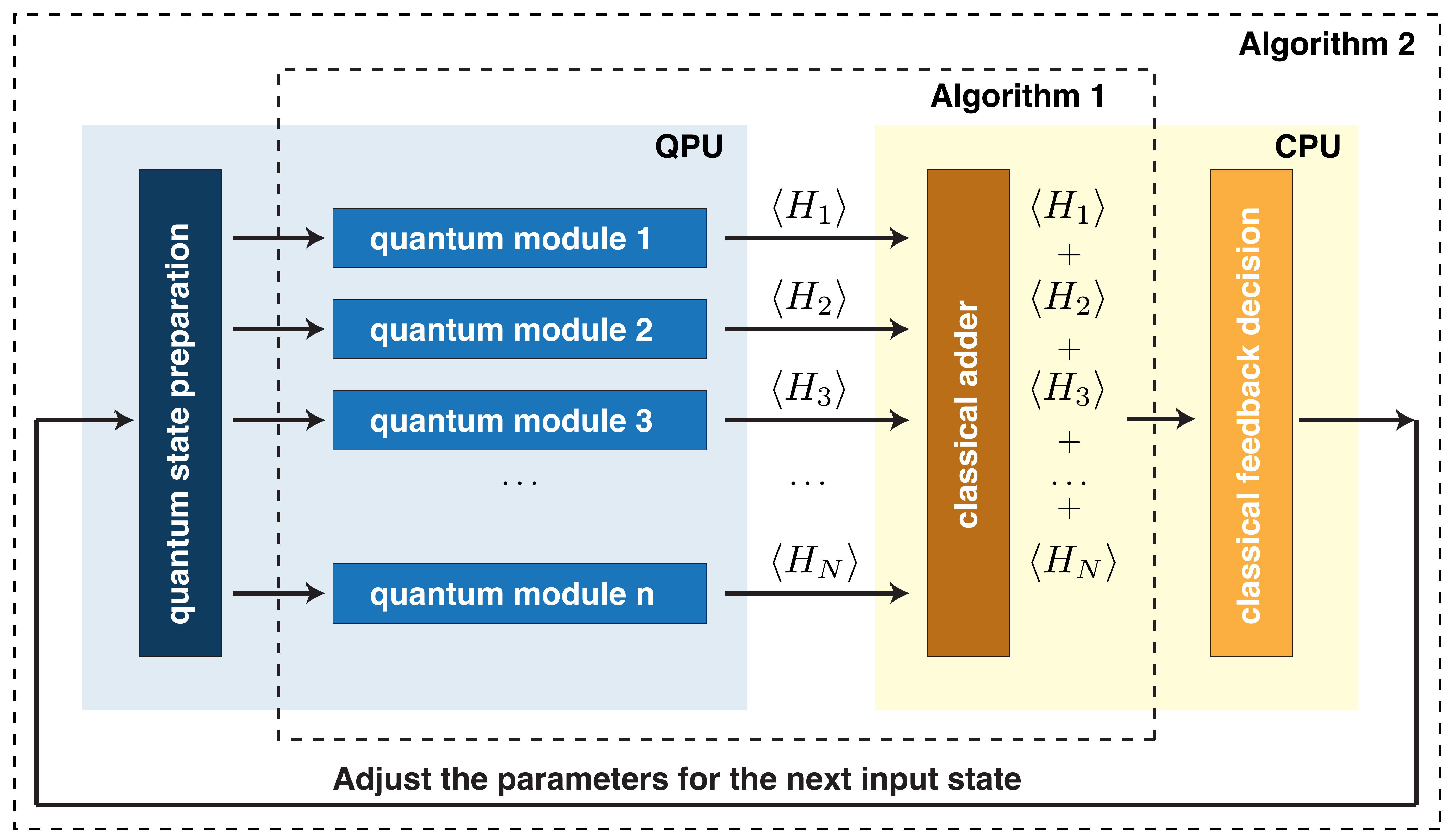}
\caption{Architecture of the quantum-variational eigensolver. 
{\bf Algorithm 1}: Quantum states that have been previously prepared, are fed into the quantum modules which compute $\avg{\h_i}$, where $\h_i$ is any given term in the sum defining $\h$. The results are passed to the CPU which computes $\avg{\h}$.
{\bf Algorithm 2}: The classical minimization algorithm, run on the CPU, takes $\avg{\h}$ and determines the new state parameters, which are then fed back to the QPU.}
\label{optimizer}
\end{figure*}

\noindent\textbf{Algorithm 1: Quantum expectation estimation\\} 
This algorithm computes the expectation value of a given Hamiltonian $\h$ for an input state $\ket{\psi}$. 
Any Hamiltonian may be written as
\be \h = \sum_{i\alpha} h^i_\alpha \sigma_\alpha^i + \sum_{ij\alpha\beta}h^{ij}_{\alpha \beta} \sigma_\alpha^i \sigma_\beta^j
+ ... 
\label{eq:hermitian_H}
\ee
for real $h$ where Roman indices identify the subspace on which the operator acts, and Greek indices identify the Pauli operator, e.g. $\alpha = x$. 
By exploiting the linearity of quantum observables, it follows that
\be \avg{\h} = \sum_{i\alpha} h^i_\alpha  \avg{\sigma_\alpha^i} + \sum_{ij\alpha\beta} h^{ij}_{\alpha \beta} \avg{ \sigma_\alpha^i \sigma_\beta^j } + ... \ee
We consider Hamiltonians that can be written as a number of terms which is polynomial in the size of the system. This class of Hamiltonians encompasses a wide range of physical systems, including the electronic structure Hamiltonian of quantum chemistry, the quantum Ising Model, the Heisenberg Model~\cite{Lloyd:2002,Ma:2011}, matrices that are well approximated as a sum of $n$-fold tensor products~\cite{Oseledets:2010,Ortiz:2001}, and more generally any $k-$sparse Hamiltonian without evident tensor product structure (see \textit{Appendix} for details). 
Thus the evaluation of $\avg \h$ reduces to the sum of a polynomial number of expectation values 
of simple Pauli operators for a quantum state $\ket{\psi}$, multiplied by some real constants. A quantum device can efficiently evaluate the expectation value of a tensor product of an arbitrary number of simple Pauli operators~\cite{Ortiz:2001}, therefore with an $n$-qubit state we can efficiently evaluate the expectation value of this $2^n \times 2^n$ Hamiltonian. 

One might attempt this using a classical computer by separately optimizing all reduced states corresponding to the desired terms in the Hamiltonian, but this would suffer from the $N$-representability problem, which is known to be intractable for both classical and quantum computers (it is in the quantum complexity class QMA-Hard~\cite{Liu:2007}).  
The power of our approach derives from the fact that quantum hardware can store a global quantum state with exponentially fewer resources than required by classical hardware, and as a result the N-representability problem does not arise.

As the expectation value of a tensor product of an arbitrary number of Pauli operators can be measured in constant time and the 
spectrum of each of these operators is bounded, to obtain an estimate with precision $p$, our approach incurs a cost of $O(|h|^2/p^2)$ repetitions.   
Thus the total cost of computing the expectation value of a state $\ket{\psi}$
is given by
$O(|h_{max}|^2 M/p^2)$, where $M$ is the number of terms in the decomposition of the Hamiltonian. The advantage of this approach is that the coherence time to make a single measurement after preparing the state is $O(1)$.  In essence, we dramatically reduce the coherence time requirement while maintaining an exponential advantage over the classical case, by adding a polynomial number of repetitions with respect to QPE.

\begin{figure*}[t]
\centering
\includegraphics[width = 14cm]{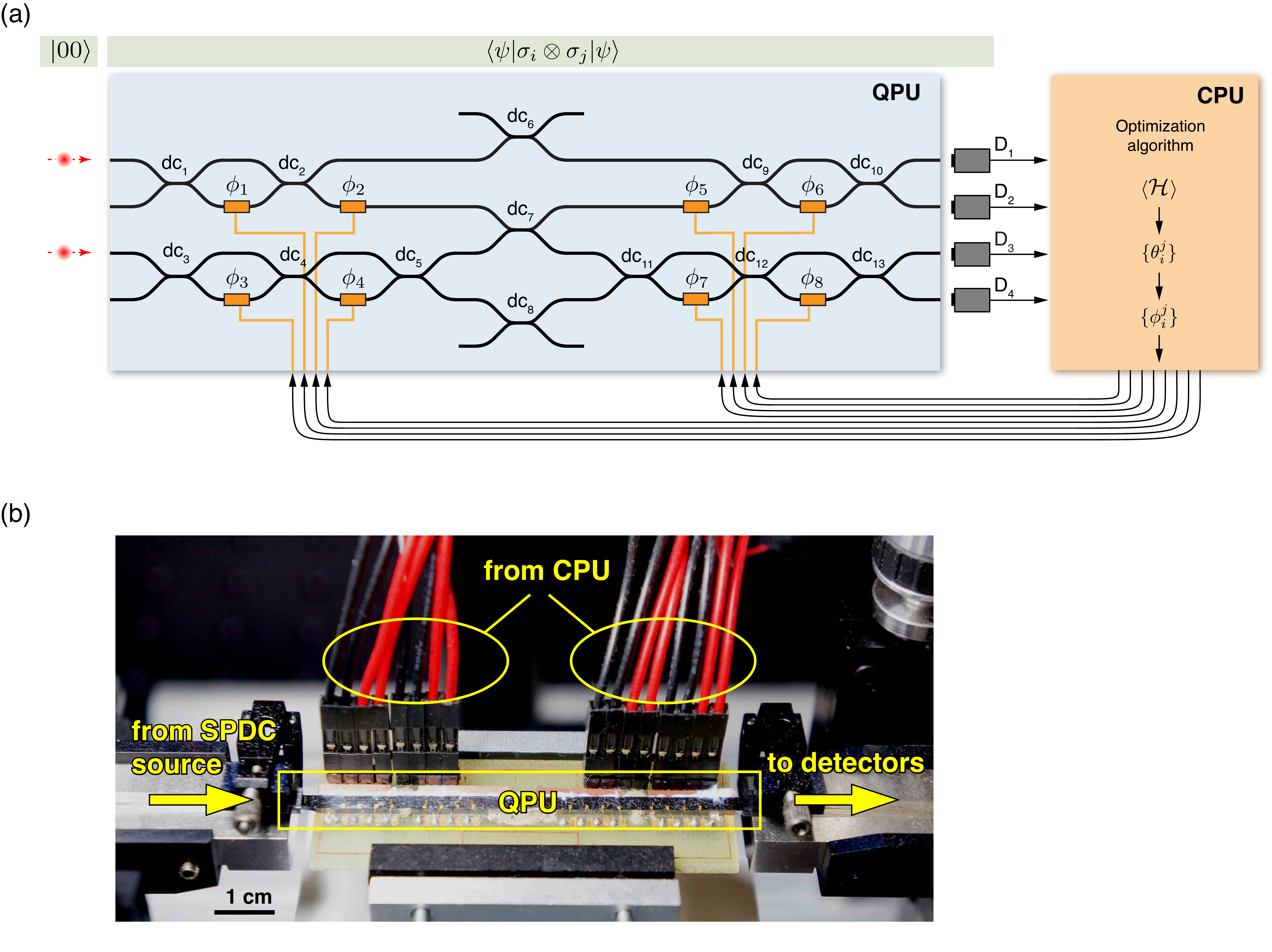}
\caption{{ Experimental implementation of our scheme.} (a) Quantum state preparation and measurement of the expectation values $\langle \psi | \sigma_i \otimes \sigma_j | \psi \rangle$ are performed using a quantum photonic chip. Photon pairs, generated using spontaneous parametric down-conversion are injected into the waveguides 
encoding the $\ket{00}$ state. The state $\ket{\psi}$ is prepared using thermal phase shifters $\phi_{1-8}$ (orange rectangles) and one CNOT gate and measured 
using photon detectors. Coincidence count rates from the detectors $\mbox{D}_{1-4}$ are passed to the CPU running the optimization algorithm. This computes the set of parameters for the next state and writes them to the quantum device. (b) A photograph of the QPU.}

\label{chip}
\end{figure*}

\noindent\textbf{Algorithm 2: Quantum variational eigensolver\\} 
The procedure outlined above replaces the long coherent evolution required by QPE by many short coherent evolutions. In both QPE and Algorithm 1 we require a good approximation to the ground state wavefunction to compute the ground state eigenvalue and we now consider this problem. Previous approaches have proposed to prepare ground states by adiabatic evolution~\cite{Aspuru:2005}, or by the quantum metropolis algorithm~\cite{Yung03012012}. Unfortunately both of these require long coherent evolution. 
Algorithm 2 is a variational method to prepare the eigenstate and, by exploiting Algorithm 1, requires short coherent evolution. Algorithm 1 and 2 and their relationship are shown in Fig. \ref{optimizer} and detailed in the \textit{Appendix}. 

It is well known that the eigenvalue problem for an observable represented by an operator $\h$ can be restated as a variational problem on the
Rayleigh-Ritz quotient~\cite{Rayleigh:1870,Ritz:1908}, such that the eigenvector $\ket{\psi}$ corresponding to the lowest eigenvalue is the $\ket{\psi}$ that minimizes
\be \frac{\bra{\psi}\h\ket{\psi}}{\braket{\psi}{\psi}} .\ee
By varying the experimental parameters in the preparation of $\ket{\psi}$ and computing the Rayleigh-Ritz quotient using Algorithm 1 as a subroutine in a classical minimization, one may prepare unknown eigenvectors.  At the termination of the algorithm, a simple prescription for the reconstruction of the eigenvector is stored in the final set of experimental parameters that define $\ket{\psi}$.

\begin{figure}[t!]
    \centering
    \includegraphics[width = 8.5cm]{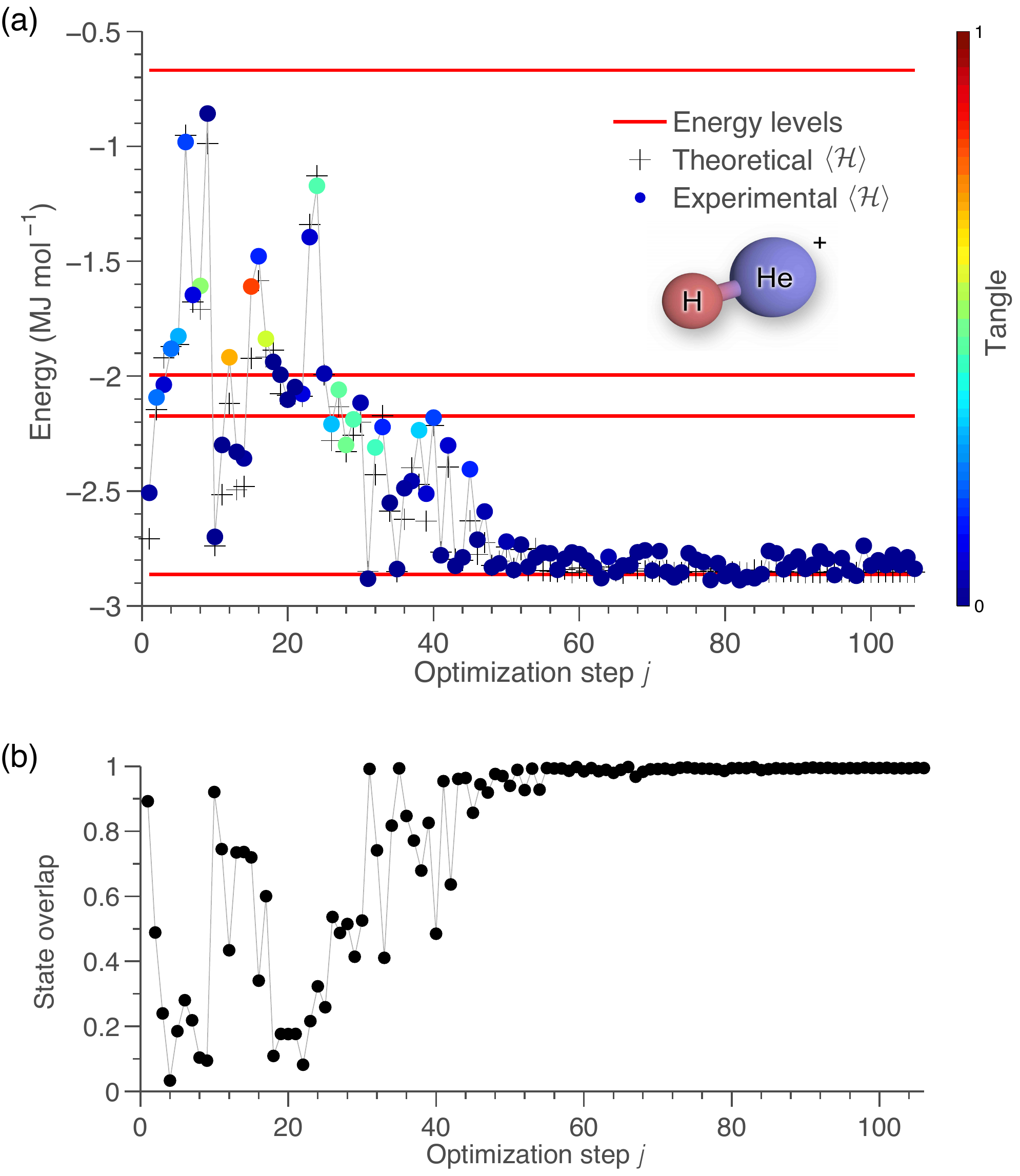}\\
 \caption{{Finding the ground state of $\mbox{He-H}^+$ for a specific molecular separation, $R=90$~pm.} (a) Experimentally computed energy $\avg{\h}$ (colored dots) as a function of the optimization step $j$. The color represents the tangle (degree of entanglement) of the physical state, estimated directly from the state parameters $\{\phi_i^j\}$. The red lines indicate the energy levels of $\h(R)$. The optimization algorithm clearly converges to the ground state of the molecule, which has small but non zero tangle. The crosses show the energy calculated at each experimental step, assuming an ideal quantum device. (b) Overlap $|\avg{\psi^{j} | \psi^{G}}|$ between the experimentally computed state $\ket{\psi^{j}}$ at each the optimization step $j$ and the 
theoretical ground state of $\h$, $\ket{\psi^G}$. Further details are provided in the \textit{Appendix}.}
\label{optimization}
\end{figure}


If a quantum state is characterized by an exponentially large number of parameters, it cannot be prepared with a polynomial number of operations. The set of efficiently preparable states are therefore characterized by polynomially many parameters, and we choose a particular set of ansatz states of this type.
Under these conditions, a classical search algorithm on the experimental parameters which define 
$\ket{\psi}$, needs only explore a polynomial number of dimensions---a requirement for the search to be efficient. 

One example of a quantum state parametrized by a polynomial number of parameters is the unitary coupled cluster
ansatz~\cite{Bartlett:2006}
\begin{equation}
  \ket{\Psi} = e^{T - T^\dagger} \ket{\Phi}_{ref}
  \label{ucc}
\end{equation}
where $T$ is the cluster operator (defined in the \textit{Appendix}) and $\ket{\Phi}_{ref}$ is some reference state,
normally taken to be the Hartree-Fock ground state. There is currently no known efficient classical algorithm based on these ansatz states. However, non-unitary coupled cluster ansatz is sometimes referred to as the ``gold standard of quantum chemistry'' as it is the standard of accuracy to which other methods in quantum chemistry are often compared.  The unitary version of this ansatz is thought to yield superior results to even this ``gold standard''~\cite{Bartlett:2006}. Details of efficient construction of the unitary coupled cluster state using a quantum device are given in the \textit{Appendix} (see also Ref.~\cite{Yung:2013}).

\noindent\textbf{Prototype demonstration\\}
We have implemented the QPU 
using integrated quantum photonics technology~\cite{Obrien:2009eu}. Our device, shown schematically in Fig.~\ref{chip} is a reconfigurable waveguide chip that implements several single qubit rotations and one two-qubit entangling gate and can prepare an arbitrary two-qubit pure state. This device operates across the full space of possible configurations with mean statistical fidelity $F > 99\%$~\cite{Shadbolt:2011bw}. 
The state is prepared, and measured in the Pauli basis, by setting 8 voltage driven phase shifters and counting photon detection events with silicon single photon detectors.

The ability to prepare an arbitrary two-qubit separable or entangled state enables us to investigate $4\times4$ Hamiltonians.  For the experimental demonstration of our algorithm 
we choose a problem from quantum chemistry, namely determining the bond dissociation curve of the molecule He-H$^+$ 
in a minimal basis. 
\begin{figure}[t!]
    \centering
    \includegraphics[width = 8.5cm]{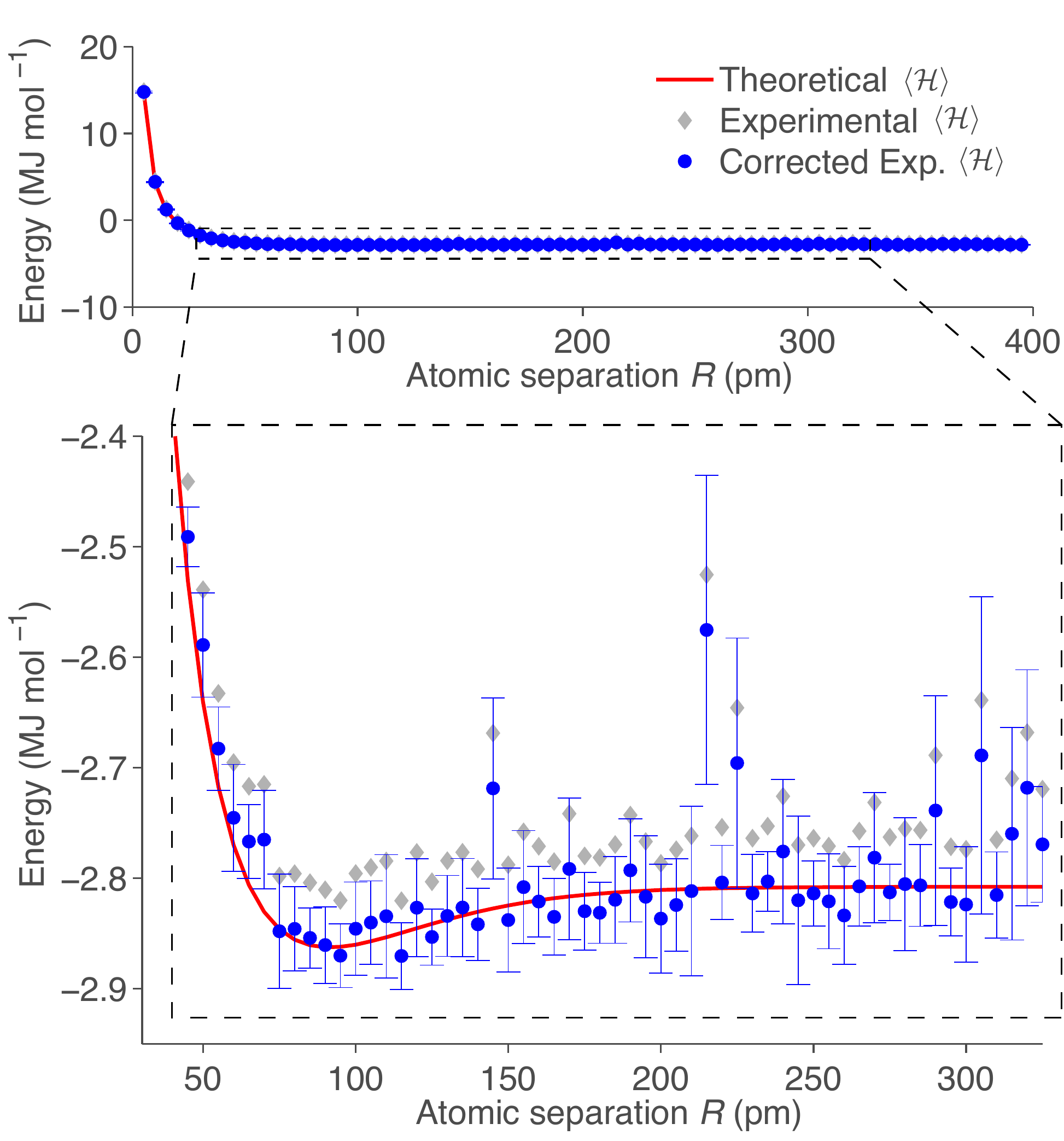}
\caption{{Bond dissociation curve of the $\mbox{He-H}^+$ molecule.} This curve is obtained by repeated computation of the ground state energy (as shown in Fig.~\ref{optimization}) for several $\h(R)$. 
The magnified plot shows that after correction for the measured systematic error, the data overlap with the theoretical energy curve and importantly we can resolve the molecular separation of minimal energy. Error bars show the standard deviation of the computed energy.}
\label{dissociation}
\end{figure}
The full configuration interaction Hamiltonian for this system has dimension 4, and can be written compactly as
\be\h(R) = \sum_{i\alpha} h^i_\alpha(R) \sigma_\alpha^i + \sum_{ij\alpha\beta}h^{ij}_{\alpha \beta}(R) \sigma_\alpha^i \sigma_\beta^j\ee
The coefficients $h^i_\alpha(R)$ and $h^{ij}_{\alpha \beta}(R)$ were determined using the PSI3 computational package~\cite{PSI3} and tabulated in the \textit{Appendix}.

In order to compute the bond dissociation of the molecule, we use Algorithm 2 to compute its ground state for a range of values of the nuclear separation $R$. In Fig.~\ref{optimization} we report a representative optimization run for a particular nuclear separation, demonstrating the convergence of our algorithm to the ground state of $\h(R)$ in the presence of experimental noise. Fig.~\ref{optimization}(a) demonstrates the convergence of the average energy, while Fig.~\ref{optimization}(b) demonstrates the convergence of the overlap $|\avg{\psi^{j} | \psi^{G}}|$ of the current state $\ket{\psi^{j}}$ with the target state $\ket{\psi^{G}}$. 
The color of each entry in Fig.~\ref{optimization}(a) represents the tangle (absolute concurrence squared) of the state at that step of the algorithm.  
It is known that the volume of separable states is doubly-exponentially small with respect to the rest of state space~\cite{Szarek:2005}.
Thus, the ability to traverse non-separable state space increases the number of paths by which the algorithm can converge and will be a requirement for future large-scale implementations.
Moreover, it is clear that the ability to produce entangled states is a necessity for the accurate description of general quantum systems where eigenstates may be non-separable, for example the ground state of the He-H$^+$ Hamiltonian has small but not negligible tangle.

Repeating this procedure for several values of $R$, we obtain the bond dissociation curve which is reported in Fig.~\ref{dissociation}. 
This allows for the determination of the equilibrium bond length of the molecule, which was found to be 
$R$=92.3$\pm$0.1~pm with a corresponding ground state electronic energy of $E$= -2.865$\pm$0.008 MJ/mol. This energy has been corrected for experimental error using a method fully described in the \textit{Appendix}. 
The corresponding theoretical curve shows the numerically exact energy derived from a full configuration interaction calculation of the molecular system in the same basis. More than 96\% of the experimental data are within chemical accuracy with respect to the theoretical values. 
At the conclusion of the optimization, we retain full knowledge of the experimental parameters, which can be used for efficient reconstruction of the state $\ket{\psi}$ in the event that additional physical or chemical properties are required.

\noindent\textbf{Discussion\\} 
Algorithm 1 uses relatively few quantum resources compared to QPE. Broadly speaking, QPE requires a large number of $n$-qubit quantum controlled operations to be performed in series---placing considerable demands on the number of components and coherence time---while the inherent parallelism of our scheme enables a small number of $n$-qubit gates to be exploited many times, drastically reducing these demands.
Moreover, adding control to arbitrary unitary operations in practice is difficult if not impossible for current quantum architectures 
(although a proposed scheme to add control to arbitrary unitary operations has recently been demonstrated~\cite{Zhou:2011}). To give a numerical example, the QPE circuit for a 4 x 4 Hamiltonian such as that demonstrated here would require at least 12 CNOT gates, while our method only requires one. 

In implementing Algorithm 2, the device prepares ansatz states that are defined by a polynomial set of parameters. This ansatz might be chosen based on knowledge of the physical system of interest (as for the unitary coupled cluster and typical quantum chemistry ans\"atze) thus determining the device design. However, our architecture allows for an alternative, and potentially more promising approach, where the device is first constructed based on the available resources and we define the set of states that the device can prepare as the ``device ansatz''. Due to the quantum nature of the device, this ansatz can be very distinct from those used in traditional quantum chemistry. With this alternative approach the physical implementation is then given by a known sequence of quantum operations with adjustable parameters---determined at the construction of the device---with a maximum depth fixed by the coherence time of the physical qubits. This approach, while approximate, provides a variationally optimal solution for the given quantum resources and may still be able to provide qualitatively correct solutions, just as approximate methods do in traditional quantum chemistry (for example Hartree Fock). 
The unitary coupled cluster ansatz (Eq.~\ref{ucc}) provides a concrete example where our approach provides an exponential advantage over known classical techniques.

We have developed and experimentally implemented a new approach to solving the eigenvalue problem with quantum hardware. 
Algorithm 1 shares with QPE the need to prepare a good approximation to the ground state, but replaces a single long coherent evolution by a number of shorter coherent calculations proportional to the number of terms in the Hamiltonian. While the effect of errors on each of these calculations is the same as in QPE, the reliance on a number of separate calculations makes the algorithm sensitive to variations in state preparation between the separate quantum calculations. This effect requires further investigation. 
In Algorithm 2, we experimentally implemented a ground state preparation procedure through a direct variational algorithm on the control parameters of the quantum hardware. 
Larger calculations will require a choice of ansatz, for which there are two possibilities. One could experimentally implement chemically motivated ans\"atze such as the unitary coupled cluster method described in the \textit{Appendix}. Alternatively one could pursue those ans\"atze that are most easy to implement experimentally---creating a new set of device ans\"atze states which would require classification in terms of their overlap with chemical ground states.  Such a classification would be a good way to determine the value of a given experimental advance---for ground state problems it is best to focus limited experimental resources on those efforts that will most enhance the overlap of preparable states with chemical ground states. In addition to the above issues, which we leave to future work, an interesting avenue of research is to ask whether the conceptual approach described here could be used to address other intractable problems with quantum-enhanced computation. Examples that can be mapped to the ground state problem, and where the n-representability problem does not occur, include search engine optimisation and image recognition. 
It should be noted that the approach presented here requires no control or auxiliary qubits, relying only on measurement techniques that are already well established. For example, in the two qubit case, these measurements are identical to those performed in Bell inequality experiments.

Quantum simulators with only a few tens of qubits are expected to outperform the capabilities of conventional computers, not including open questions regarding fault tolerance and errors/precision. Our scheme would allow such devices to be implemented using dramatically less resources than the current best known approach.

\clearpage

 \noindent\textbf{APPENDIX\\}

 \noindent\textbf{SUPPLEMETARY THEORY\\}
 \noindent\textbf{Quantum eigenvector preparation algorithm\\}
 
 Below we detail the steps involved in implementing Algorithm 2.
 
\begin{enumerate}
    \item Design a quantum circuit, controlled by a set of experimental parameters $\{\theta_i\}$, which can  
   prepare a class of states. 
   Using this device, prepare the initial state $\ket{\psi^0}$ and define the objective function 
   $f (\{ \theta_i^n\}) =   \bra{\psi(\{ \theta_i^n\})}\h\ket{\psi(\{ \theta_i^n\})}$, which efficiently maps the set of experimental parameters to the expectation value of the Hamiltonian and is computed in this work by Algorithm 1.  $n$ denotes the current iteration of the algorithm.
  \item Let $n=0$
  \item Repeat until optimization is completed \\
  \begin{enumerate}
    \item Call Algorithm 1 with  $\{\theta_i\}$ as input:
    \begin{enumerate}
    \item Using the QPU, compute $\avg{\sigma_\alpha^i}$, $\avg{ \sigma_\alpha^i \sigma_\beta^j }$,
          $\avg{\sigma_\alpha^i \sigma_\beta^j \sigma_\gamma^k }$, $...$, on $\ket{\psi^n}$ for all terms of $\h$.
    \item Classically sum on CPU the values from the QPU with their appropriate weights, $h$, to obtain $f(\{ \theta_i^n \})$
    \end{enumerate}
    \item Feed $f(\{ \theta_i^n \})$ to the classical minimization algorithm (e.g. gradient descent or Nelder-Mead Simplex method),
          and allow it to determine $\{ \theta_i^{n+1} \}$.
    \end{enumerate}
\end{enumerate}

\vspace{5mm}  

\noindent\textbf{Second Quantized Hamiltonian}\\
When taken with the Born-Oppenheimer approximation, the Hamiltonian of
an electronic system can be generally written~\cite{Helgaker2002} as 
\begin{equation}
  \h(R) = \sum_{pq} h_{pq}(R) \hat a_p^\dagger \hat a_q + \sum_{pqrs}h_{pqrs}(R) \hat a_p^\dagger \hat a_q^\dagger \hat a_r \hat a_s
\end{equation}
where $\hat a_i^\dagger$ and $\hat a_j$ are the fermionic creation and annihilation operators
that act on the single particle basis functions chosen to represent the electronic system and obey
the canonical anti-commutation relations $\{\hat a_i^\dagger, \hat a_j\} = \delta_{ij}$ and
 $\{\hat a_i, \hat a_j\} = \{\hat a_i^\dagger, \hat a_j^\dagger \} = 0$. $R$ is a vector
representing the positions of the Nuclei in the system, and is fixed for any given geometry.
The constants $h_{pq}(R)$ and $h_{pqrs}(R)$ are evaluated using an initial Hartree-Fock calculation and relate the second
quantized Hamiltonian to the first quantized Hamiltonian.  They are calculated as
\begin{align}
 h_{pq} &= \int dr \  \chi_p(r)^* \left( -\frac{1}{2} \nabla^2 - \sum_\alpha \frac{Z_\alpha}{|r_\alpha - r|} \right) \chi_q(r) \\
 h_{pqrs} &= \int dr_1 \ dr_2 \ \frac{\chi_p(r_1)^* \chi_q(r_2)^* \chi_r(r_1) \chi_s(r_2)}{|r_1 - r_2|} 
\end{align}
where $\chi_p(r)$ are single particle spin orbitals, $Z_\alpha$ is the nuclear charge, and $r_\alpha$
is the nuclear position. From the definition of the Hamiltonian, it is clear that the number of terms in the Hamiltonian
is $O(N^4)$ in general, where $N$ is the number of single particle basis functions used. The
map from the Fermionic algebra of the second quantized Hamiltonian to the distinguishable
spin algebra of qubits is given by the Jordan-Wigner transformation~\cite{Jordan:1928},
which for our purposes can be concisely written as
\begin{align}
 \hat a_j \rightarrow I^{\otimes j-1} \otimes \sigma_+ \otimes \sigma_z^{\otimes N - j} \\
 \hat a_j^\dagger \rightarrow I^{\otimes j-1} \otimes \sigma_- \otimes \sigma_z^{\otimes N - j}
\end{align}
where $\sigma_+$ and $\sigma_-$ are the Pauli spin raising and lowering operators respectively.
It is clear that this transformation does not increase the number of terms present in the
Hamiltonian, it merely changes their form and the spaces on which they act.  Thus the
requirement that the Hamiltonian is a sum of polynomially many products of Pauli operators
is satisfied.  As a result, the expectation value of any second quantized chemistry Hamiltonian
can be efficiently measured with our scheme.

For the specific case of He-H$^+$ in a minimal, STO-3G basis, it turns out that full
configuration interaction (FCI) Hamiltonian has dimension four, thus a more compact
representation is possible than in the general case.  In this case, the FCI
Hamiltonian can be written down for each geometry expanded in terms of the tensor products
of two Pauli operators.  Thus the Hamiltonian is given explicitly by an FCI calculation
in the PSI3 computational package~\cite{PSI3} and can be written as
\begin{equation}
\h(R) = \sum_{i\alpha} h^i_\alpha(R) \sigma_\alpha^i + \sum_{ij\alpha\beta} h^{ij}_{\alpha \beta}(R) \sigma_\alpha^i \sigma_\beta^j
\end{equation}
 
\vspace{5mm}  

\noindent\textbf{Unitary Coupled Cluster Theory}\\
One example 
of a state which is efficiently preparable
on a quantum computer, but not so on a classical computer is the unitary
coupled cluster expansion~\cite{Bartlett:2006}.  The unitary coupled cluster theory method is a variational
ansatz which takes the form
\begin{equation}
  \ket{\Psi} = e^{T - T^\dagger} \ket{\Phi}_{ref}
\end{equation}
where $\ket{\Phi}_{ref}$ is some reference state, usually the Hartree Fock ground state, and
$T$ is the cluster operator for an N electron system defined by
\begin{equation}
 T = T_1 + T_2 + T_3 + ... + T_N
\end{equation}
with
\begin{align}
T_1 &= \sum_{pr} t_p^r \hat a^\dagger_p \hat a_r \\
T_2 &= \sum_{pqrs} t_{pq}^{rs} \hat a^\dagger_p \hat a^\dagger_q \hat a_r \hat a_s
\end{align}
where repeated indices imply summation as in the main text, and higher order terms follow logically.  It is clear that by construction
the operator $(T - T^\dagger)$ is anti-hermitian, and exponentiation
maps it to a unitary operator $U=e^{(T - T^\dagger)}$. For any fixed excitation level $k$, 
the reduced cluster operator is written as
\begin{equation}
  T^{(k)} = \sum_{i=1}^k T_i
\end{equation}
Unfortunately, in general no efficient implementation of this ansatz has yet been developed for a 
classical computer, even for low order cluster operators due to the non-truncation of the BCH series~\cite{Bartlett:2006}.
The reduced anti-hermitian cluster operator $(T^{(k)}-T^{(k)\dagger})$ is
the sum of a polynomial number of terms in the number of one electron basis
functions, namely it contains a number of terms $O(N^k(M-N)^k)$ where M is the number of single
particle orbitals.  By defining an effective Hermitian Hamiltonian $\h=i(T^{(k)}-T^{(k)\dagger})$
and performing the Jordan-Wigner transformation to reach a Hamiltonian that
acts on the space of qubits, $\tilde \h$, we are left with a Hamiltonian which is a sum
of polynomially many products of Pauli operators.  The problem then reduces to the
quantum simulation of this effective Hamiltonian, $\tilde \h$, which can be done in polynomial
time using the procedure outlined by Ortiz et al.~\cite{Ortiz:2001}.
This represents one example of a state which can be efficiently prepared on a quantum device, which cannot be efficiently prepared by any known means on a classical computer.

\vspace{5mm}  

\noindent\textbf{Finding excited states}\\
Frequently, one may be interested in eigenvectors and eigenvalues related to excited states (interior eigenvalues).
Fortunately our scheme can be used with only minor modification to find these excited states by repeating the procedure  
on $\h_{\lambda} = (\h - \lambda)^2$.  The folded spectrum method~\cite{MacDonald:1934,Wang:1994}
allows a variational method to converge to the eigenvector closest to the shift parameter $\lambda$.  By scanning through
a range of $\lambda$ values, one can recover the eigenvectors and eigenvalues of interest.  Although this
operation incurs a small polynomial overhead ---the number of terms in the Hamiltonian is quadratic with respect to the original Hamiltonian---  this extra cost is marginal compared to the cost of solving the problem classically.

\vspace{5mm}  

\noindent\textbf{Application to $k-$sparse Hamiltonians}\\
The method described in the main body of this work may be applied to general $k-$sparse Hamiltonian
matrices which are row-computable even when no efficient tensor decomposition is evident with only minor modification.  A Hamiltonian $\h$ is referred to as $k-$sparse if there are at most $k$ non-zero elements in each row and column of the matrix and row computable if there is an efficient algorthim for finding the locations and values of the non-zero matrix elements in each row of $\h$.

Let $\h$ be a $2^n \times 2^n$ $k-$sparse row-computable Hamiltonian.  A result by Berry et al.~\cite{Berry:2007} shows that $\h$ may be decomposed as $\h = \sum_{j=1}^m \h_j$ with $m=6k^2$, $\h_j$ being a $1-$sparse matrix and each $\h_j$ may be efficiently simulated ($e^{-i \h_j t}$ may be acted on the qubits) by making only $O(\log^* n)$ queries to the Hamiltonian $\h$. 
Alternatively, a more recent result by Childs et al.~\cite{Childs:2011} has found that it possible to use a star decomposition of the Hamiltonian such that $m=6k$ and each $\h_j$ is now a galaxy which can be efficiently simulated using $O(k + \log^*N)$ queries to the Hamiltonian. Either of these schemes may be used to implement our algorithm efficiently for a general $k-$sparse matrix, and the choice may be allowed to depend on the particular setup available. 
Following a prescription by Knill et al.~\cite{Knill:2007}, the ability to simulate $\h_j$ is sufficient for efficient measurement of the expectation value $\langle \h_j \rangle$.  After determining these values, one may proceed as before in the algorithm as outlined in the main text and use them to determine new parameters for the classical minimization.

\vspace{5mm}  

\noindent\textbf{Classical optimization algorithm}\\
For the classical optimization step of our integrated processor we implemented the Nelder-Mead (NM) algorithm~\cite{Nelder:1965}, a simplex-based direct search (DS) method for unconstrained minimization of objective functions. Although in general NM can fail because of the deterioration of the simplex geometry or lack of sufficient decrease, the convergence of this method can be greatly improved by adopting a restarting strategy.
Although other DS methods, such as the gradient descent, can perform better for smooth functions, these are not robust to the noise which makes the objective function non-smooth under experimental conditions. NM has the ability to explore neighboring valleys with better local optima and likewise this exploring feature usually allows NM to overcome non-smoothnesses. 
We verified that the gradient descent minimization algorithm is not able to converge to the ground state of our Hamiltonian under the experimental conditions, mainly due to the poissonian nature of our photon source and the accidental counts of the detection system, while NM converged to the global minimum in most optimization runs. 
\vspace{5mm}  

\noindent\textbf{Computational Scaling}\\
In this section, we demonstrate the polynomial scaling of each iteration of our algorithm with respect to system size, and contrast that with the exponential scaling of the current best-known classical algorithm for the same task.  Suppose that the algorithm has progressed to an iteration $j$ in which we have prepared a state vector $\ket{\psi^j}$ which is stored in $n$ qubits and parameterized by the set of parameters $\{\theta^j_i\}$.  

We wish to find the average value of the Hamiltonian, $\avg{\h}$ on this state.  We will assume that there are $M$ terms comprising the Hamiltonian, and assume that $M$ is polynomial in the size of the physical system of interest.  Without loss of generality, we select a single term 
from the Hamiltonian, $\h_i$ that acts on $k$ bits of the state, and denote the average of this term by $\avg{\h_i} = h \avg{\tilde \sigma}$ where $h$ is a constant 
and $\tilde \sigma$ is the $k-$fold tensor product of Pauli operators acting on the system.  As the expectation value of a tensor product of an arbitrary number of
Pauli operators can be measured in constant time and the spectrum of 
each of these operators is bounded, if the desired precision on the value is given by $p$, we expect the cost of this estimation to be $O(|h|^2/p^2)$ repetitions of the preparation and measurement procedure.  Thus we estimate the cost of each function evaluation to be
$O(|h_{max}|^2 M/p^2)$.  For most modern classical minimization algorithms (including the Nelder-Mead simplex method~\cite{Nelder:1965}), the cost of a single update
step, scales linearly or at worst polynomially in the number of parameters included in the minimization~\cite{Fletcher:1987}.
By assumption, the number of parameters in the set $\{\theta^j_i\}$, is polynomial in the system size.   Thus the total cost per iteration is roughly given by $O(n^r |h_{max}|^2 M / p^2)$ for some small constant $r$ which is determined by the encoding of the quantum state and the 
classical minimization method used.

Contrasting this to the situation where the entire algorithm is performed classically, a much different scaling results.  Storage
of the quantum state vector $\ket{\psi^j}$ using currently known exact encodings of quantum states, requires knowing
$2^n$ complex numbers.  Moreover, given this quantum state, the computation of the expectation value 
$\avg{\tilde \sigma} = \bra{\psi^j} \tilde \sigma \ket{\psi^j}$ using modern methods requires $O(2^n)$ floating point operations.  Thus a single function evaluation is expected to require exponential resources in both storage and computation when performed on a classical computer.  Moreover, the number of parameters which a classical minimization algorithm must manipulate to represent this state exactly is $2^n$.  Thus performing even a single minimization step to determine $\ket{\psi^{j+1}}$ requires an exponential number of function evaluations, each of which carries an exponential cost.  One can roughly estimate the scaling of this procedure as $O(M2^{n(r+1)})$

From this coarse analysis, we conclude that our algorithm attains an exponential advantage in the cost of a single iteration over
the best known classical algorithms, provided the assumptions on the Hamiltonian and quantum state are satisfied.  While
convergence to the final ground state must still respect the known complexity QMA-Complete complexity of this task~\cite{Kempe:2006}, we believe this still demonstrates the value of our algorithm, especially in light of the limited quantum resource requirements.
\vspace{5mm}  
 
\noindent\textbf{Mapping from the state parameters to the chip phases.} 
The set of phases $\{\theta_i\}$, which uniquely identifies the state $\ket{\psi}$, is not equivalent to the phases which are written to the photonic circuit $\{\phi_i\}$, since the chip phases are also used to implement the desired measurement operators $\sigma_\alpha \otimes \sigma_\beta$. Therefore, knowing the desired state parameters and measurement operator we compute the appropriate values of the chip phases on the CPU at each iteration of the optimization algorithm.  
\vspace{5mm}  

\noindent\textbf{EXPERIMENTAL DETAILS\\}
\noindent\textbf{Estimation of the error on $\avg\h$}\\
We performed measurements of the statistical and systematic errors that affect our computation of $\avg\h$.

\noindent\textbf{Statistical errors}
Statistical errors due to the Poissonian nature of single photon statistics are intrinsic to the estimation of expectation values in quantum mechanics. 
 
These errors can be arbitrarily reduced at a sublinear cost of measurement time (i.e. efficiently) since the magnitude of error is proportional to the square root of the count rate. We experimentally measured the standard deviation of an expectation value $\avg{\h_i}$ for a particular state using 50 trials. The total average coincidence rate was $\sim$1500/s. The standard deviation was found to be 37KJ/mol, which is  comparable with the error observed in the measurement of the ground state energy shown in Fig. 4

The minima of the potential energy curve was determined by a generalized least squares procedure to 
fit a quadratic curve to the experimental data points in the region $R=(80,100)$ pm, as is common in the use of trust region searches for minima~\cite{Conn:1987}, using the inverse experimentally measured variances as weights.  Covariances determined by the generalized least squares procedure were used as input to a Monte Carlo sampling procedure to determine the minimum energy and equilibrium bond distance as well as their uncertainties assuming Gaussian random error.  The uncertainties reported represent standard deviations.  Sampling error in the Monte Carlo procedure was $3 \times 10^{-4}$ pm for the equilibrium bond distance and $3 \times 10^{-8}$ MJ/mol for the energy.

\noindent\textbf{Systematic errors}
In all the measurements described above we observed a constant and reproducible small shift, $\epsilon = 50KJ/mol$, of the expectation value with respect to the theoretical value of the energy. There are at least three effects which contribute to this systematic error. 

Firstly, the down-conversion source that we use in our experiment does not produce the pure two photon state that is required for high-fidelity quantum interference. In particular, higher order photon number terms and, more significantly, photon distinguishability both degrade the performance of our entangling gate and thus the preparation of the state $\ket{\psi}$. This results in a shift of the measured energy $\bra{\psi}\h\ket{\psi}$. Higher order terms could be effectively eliminated by use of true single photon sources (such as quantum dots or nitrogen vacancy centers in diamond), and there is no fundamental limit to the degree of indistinguishability which can be achieved through improved state engineering.

Secondly, imperfections in the implementation of the photonic circuit also reduce the fidelity with which $\ket{\psi}$ is prepared and measured. Small deviations from designed beamsplitter reflectivities and interferometer path lengths, as well as imperfections in the calibration of voltage-controlled phases shifters used to manipulate the state, all contribute to this effect. However, these are technological limitations  that can be greatly improved in future realizations.

Finally, unbalanced input and output coupling efficiency also results in skewed two-photon statistics, again shifting the measured expectation value of $\avg\h$.

Another systematic effect that can be noted in Fig. 4 is that the magnitude of the error on the experimental estimation of the ground state energy increases with $R$. This is due to the fact that as $R$ increases, the first and second excited eigenstates of this Hamiltonian become degenerate, resulting in increased difficulty for the classical minimization, generating mixtures of states that increases the overall variance of the estimation.
\vspace{5mm}

\noindent\textbf{Count rate}\\
In our experiment the mean count rate, which directly determines the statistical error, was approximately 2000-4000 twofold events per second. For example, for the bond dissociation curve we measured about 100 points per optimization run. The expectation value of a given Hamiltonian was reconstructed at each point from four two-qubit Pauli measurements. In the full dissociation curve we found the ground states of 79 Hamiltonians. Hence the full experiment was performed in about 158 hours. 

State preparation is relatively fast, requiring a few milliseconds to set the phases on the chip. However 17 seconds are required for cooling the chip, resulting in a duty-cycle of $\sim 5\%$. The purpose of this is to overcome instability of the fibre-to-chip coupling due to thermal expansion of the chip during operation. This will not be an issue in future implementations where fibres will be permanently fixed to the chip's facets. Moreover the thermal phase shifters used here will also likely be replaced by alternative technologies based on the electro-optic effect.

Brighter single photon sources will considerably reduce the measurement time.

\end{document}